\documentclass[aps,prl,twocolumn,showpacs,superscriptaddress]{revtex4}

\usepackage{graphicx}
\usepackage{dcolumn}
\usepackage{bm}

\begin{document}


\title{Quantitative complementarity relations in bipartite systems}
\author{Matthias Jakob}
\affiliation{Department of Physics, Hunter College, City University 
of New York, 695 Park Avenue, New York, NY 10021  \nolinebreak}
\affiliation{Physics Department, Royal Institute of Technology (KTH), \nolinebreak
AlbaNova, Roslagstullsbacken 21, SE-10691 Stockholm, Sweden \nolinebreak}
\author{J\'anos A.\ Bergou}
\affiliation{Department of Physics, Hunter College, City University 
of New York, 695 Park Avenue, New York, NY 10021  \nolinebreak}

\date{\today}

\begin{abstract}
We introduce a complete set of complementary quantities in bipartite,  
two-dimensional systems. Complementarity then relates the quantitative 
entanglement measure concurrence which is a bipartite property to the
single-particle quantum properties predictability 
and visibility, for the most general quantum state of two qubits. 
Consequently, from an interferometric point of view,  
the usual wave-particle duality relation must be extended 
to a ``triality'' relation containing, in addition, the 
quantitative entanglement measure concurrence, which has no classical
counterpart and manifests a genuine quantum aspect of bipartite
systems. A generalized duality relation, that also governs possible
violations of the Bell's inequality, arises between single- and
bipartite properties.  
\end{abstract}

\maketitle

The concept of complementarity in its full generality is summarized in
the statement that a quantum
system may posses properties which are equally real but 
mutually exclusive \cite{bohr}. Often it is associated with 
wave particle duality, the complementary aspect between propagation
and detection. Furthermore, it is commonly related to mutually 
exclusive properties of {\it single} quantum systems. 
Here, we go beyond single-partite systems and show that
entanglement, which is a genuinely quantum concept with no classical
counterpart, is a key entry in a generalized complementarity relation
for {\it bipartite} systems.  
We consider the most general quantum state formed by two qubits 
in order to perform this task and arrive at a complementarity relation
which is formed  
by three quantities, i.e.\ a ``triality relation'' is found.  
Two of the quantities generate {\it local}, single-partite realities 
which can be related to wave particle duality 
while the third quantity is found as the entanglement measure 
{\it concurrence} \cite{wootters} which generates 
an exclusive bipartite {\it nonlocal} reality. 
As one of the consequences, violations of the Bell inequality  
may be understood to be governed by this complementarity relation. 
We relate our findings to the known quantitative complementarity 
relations between distinguishability and visibility \cite{englert} 
and single- and two-particle visibility \cite{jaeger1,jaeger2} 
which intrinsically consider bipartite systems. 
With the help of the new complementarity relation, 
we demonstrate the common origin and 
equivalence of these known complementarity relations. 
 
Suppose the most general bipartite state of two qubits,
\begin{equation}
|\Theta\rangle = a|00\rangle + b|01\rangle + c|10\rangle +
 d|11\rangle\ , 
\label{eq1}
\end{equation}
where, from normalization, $a$, $b$, $c$, and $d$ satisfy
\begin{equation}
|a|^{2} + |b|^{2} +|c|^{2}+|d|^{2} = 1\ . 
\label{eq2}
\end{equation} 
The concurrence is defined as 
${\cal{C}}=|\langle\Theta|\tilde{\Theta}\rangle|$ with  
\begin{equation}
|\tilde{\Theta}\rangle = 
(\sigma_{y}\otimes\sigma_{y})|\Theta^{\ast}\rangle,  
\;\;\;\;\;\;
\sigma_{y} = \left[ 
\begin{array}{cc}
0 & -i \\
i & 0
\end{array}
\right], \label{eq3}
\end{equation}
and $|\Theta^{\ast}\rangle$ is the complex conjugate of 
$|\Theta\rangle$. 
It is related to the entanglement 
of formation, $E(\Theta)={\cal{E}}({\cal{C}}(\Theta))$
\cite{wootters,wootters1}, where  
\begin{eqnarray}
{\cal{E}}({\cal{C}}) &=&
h\left(\frac{1+\sqrt{1+{\cal{C}}^{2}}}{2}\right), 
\label{eq4} \\
h(x) &=& -x\log_{2}x-(1-x)\log_{2}(1-x). \nonumber
\end{eqnarray} 
The concurrence is a measure of entanglement on its own and
constitutes an explicit bipartite property which describes 
phase relations shared among both parties. It is intimately related to
the two-particle visibility, ${\cal{V}}_{12}$, in an interferometric
setup \cite{jaeger1,jaeger2,abouraddy,jakob1} and will represent the
first entry in our complementarity relation. 
For the case of the most general state, Eq.\ (\ref{eq1}), we find 
\begin{equation}
{\cal{C}}(\Theta) = 2\left|ad-bc\right|. 
\label{eq5}
\end{equation}

The second quantity in the complementarity relation is 
given by the coherence ${\cal{V}}$ between 
the two orthogonal qubit states, $|0\rangle$ and $|1\rangle$. 
We note that, in contrast to the concurrence, the coherence is 
a {\it single} qubit quantity. 
When we assume that the states represent two different
alternatives in an interferometer, the coherence can be quantified by
the fringe visibility, ${\cal{V}}$, of the arising interference
pattern. ${\cal{V}}$ satisfies the standard definition 
\begin{equation}
{\cal{V}} = \frac{I_{\text{max}}-I_{\text{min}}}
{I_{\text{max}}+I_{\text{min}}}, \label{eq6}
\end{equation}
where $I_{\text{max}}$ and $I_{\text{min}}$ 
denote the maximum and minimum intensity, respectively.  
In a bipartite system, we must distinguish between 
the coherences of the two subsystems. 
The coherence between the states 
$|0\rangle$ and $|1\rangle$ of subsystem $k$ ($k=1$ or $2$), assuming 
the total system is in state $|\Theta\rangle$, can 
be introduced as 
\begin{eqnarray}
{\cal{V}}_{k} &=& 
2|\langle\Theta|\sigma_{k}^{+}|\Theta\rangle|, 
\qquad
\sigma_{k}^{+} = 
\left[
\begin{array}{cc}
0 & 1\\
0 & 0
\end{array}
\right]\ . 
\nonumber
\end{eqnarray}
From this definition, using Eq.\ (\ref{eq1}), we readily find
\begin{eqnarray}
{\cal{V}}_{1} &=& 
2|ac^{\ast}+bd^{\ast}|, \qquad
{\cal{V}}_{2} = 
2|ab^{\ast}+cd^{\ast}|.  
\label{eq7} 
\end{eqnarray}
The coherence ${\cal{V}}_{k}$, as expected, is given by the
off-diagonal elements of the reduced density matrix for subsystem
$k$ that we obtain from Eq. (\ref{eq1}) by tracing over the other
subsystem.

The third and last entry in the complementarity relation 
also generates a single particle reality. 
It is the predictability, ${\cal{P}}$, which quantifies the 
{\it a priori} knowledge of whether the subsystems are in state
$|0\rangle$ or $|1\rangle$. Again, in a bipartite system, we must
distinguish between the predictabilities of the two subsystems. 
The predictability ${\cal{P}}_{k}$, for subsystem $k$ ($k=1$
or $2$), is defined as  
\begin{eqnarray}
{\cal{P}}_{k} &=& |\langle \Theta|\sigma_{z,k}|\Theta\rangle|, 
\qquad
\sigma_{z,k} = 
\left[
\begin{array}{cc}
1 & 0 \\
0 & -1
\end{array}
\right]\ . 
\nonumber 
\end{eqnarray}
Using Eq.\ (\ref{eq1}) we find
\begin{eqnarray}
{\cal{P}}_{1} &=& 
\left| (|c|^{2} + |d|^{2}) - (|a|^{2} + |b|^{2}) \right|, 
\nonumber \\
{\cal{P}}_{2} &=& 
\left| (|b|^{2} + |d|^{2}) - (|a|^{2} + |c|^{2}) \right|\ .  
\label{eq8}
\end{eqnarray}  
The predictability, ${\cal{P}}_{k}$, as expected, is given by the
difference of the diagonal elements of the reduced density matrix of
subsystem $k$.

We point out that all of the introduced quantities correspond to
observables that can be measured in experiments. These quantities
satisfy the following complementarity relation, 
\begin{equation}
{\cal{C}}^{2}+{\cal{V}}^{2}_{k}+{\cal{P}}^{2}_{k} 
= 
\left( |a|^{2} + |b|^{2} + |c|^{2} + |d|^{2} \right)^{2} \equiv 1,
\label{eq10} 
\end{equation}
which is a simple exercise to prove and constitutes the central
result of this paper. 
In Eq.\ (\ref{eq10}) we have  assumed the system 
in the most general {\it pure} bipartite state 
$|\Theta\rangle$, as given by Eq.\ (\ref{eq1}).  
This complementarity relation holds for the most general
interferometric schemes with biparticles or biphotons, as well. 
In contrast to the duality relation 
of single partite quantum systems, it contains {\it three} 
mutually exclusive but equally real quantities. 
One of the quantities, entanglement, is a genuine quantum property of
bipartite systems, while the other two correspond to single partite
properties. Combining the single partite properties into a single
entity, for $k=1$ or $2$, 
\begin{equation}
{\cal{S}}_{k}^{2}={\cal{V}}_{k}^{2}+{\cal{P}}_{k}^{2}, 
\label{eq11}
\end{equation} 
a {\it duality} relation between bipartite and single partite
properties arises, 
\begin{equation}
{\cal{C}}^{2}+{\cal{S}}_{k}^{2} =1. 
\label{eq12}
\end{equation} 
In the language of quantum information 
theory \cite{nielsen}, ${\cal{S}}_{k}$ forms a 
{\it local} quantity whose constituents 
${\cal{P}}_{k}$ and ${\cal{V}}_{k}$   
can be changed under local unitary transformations 
into ${\cal{P}}_{k}^{\prime}$ and ${\cal{V}}_{k}^{\prime}$,  
satisfying the condition ${\cal{P}}_{k}^{2}+{\cal{V}}_{k}^{2}=
({\cal{P}}_{k}^{\prime})^{2}+({\cal{V}}_{k}^{\prime})^{2}$. In
particular, ${\cal{S}}_{k}$ can be all visibility with no
predictability or, alternatively, all predictability with no
visibility. By contrast, the concurrence ${\cal{C}}$ is a {\it 
nonlocal} property which remains invariant under local unitary
transformations, i.e.\ ${\cal{C}}={\cal{C}}^{\prime}$. 

We are now in a position to compare the new complementarity relation 
to previous ones. The first complementarity relation for a large class
of pure {\it bipartite} quantum systems was derived by 
Jaeger {\it et al.} \cite{jaeger1}. 
It relates {\it two-particle visibility} 
${\cal{V}}_{12}$ to {\it single-particle visibility} 
${\cal{V}}_{k}$ ( $k=1,2$), in the form of an inequality,  
\begin{equation}
{\cal{V}}_{12}^{2}+{\cal{V}}_{k}^{2} \leq 1. 
\label{eq13}
\end{equation}
A similar relation has actually been demonstrated in a recent
experiment \cite{abouraddy1}. It is worth noting at this point that
the appearance of an inequality in a {\it pure} system, as contrasted
to the equality relations    
(\ref{eq10}) and (\ref{eq12}), is the manifestation of 
a missing entity. Furthermore, this complementarity relation has
been derived for a restricted set of transducers in the
interferometer, viz.\ each transducer was taken to consist of 
{\it symmetric} beam splitters together with phase shifters.  
 
In a subsequent publication Jeager {\it et al.} 
\cite{jaeger2} demonstrated that a stronger complementarity 
relation in form of an equality can be found  
if {\it arbitrary} local unitary and unimodular mappings 
between the subspaces, i.e.\ arbitrary transducers, 
are allowed. In terms of a generalized single particle visibility, 
$V_{k}$ ($k=1,2$), and a generalized two-particle visibility, 
$V_{12}$, this complementarity relation is given as
\begin{equation}
V_{12}^{2}+V_{k}^{2}=1\ . 
\label{eq14}
\end{equation}  
It was later recognized that the 
two-particle visibility $V_{12}$ is directly related to 
the concurrence ${\cal{C}}$ \cite{abouraddy1,jakob1}. 
Since the concurrence is a proper entanglement measure,
it is invariant under local unitary transformations. In fact, the
quantities 
$V_{12}\equiv {\cal{C}}$ and ${\cal{V}}_{12}\equiv{\cal{C}}$ 
in (\ref{eq13}) and (\ref{eq14}) are identical. This has
not been recognized in Refs.\ \cite{jaeger1,jaeger2}. 
Thus, the complementarity relation (\ref{eq14}) is a 
special and restricted case of the general 
duality relation (\ref{eq12}) between 
single- and bipartite properties. 
Indeed, as already mentioned after Eq.\ (\ref{eq12}), 
the single-partite property ${\cal{S}}_{k}$ can be transformed 
with local unitary transformations, i.e.\ proper transducers, 
into a single-particle visibility equivalent to the 
generalized single-particle visibility $V_{k}$ introduced in
\cite{jaeger2}.  
The general complementarity relation (\ref{eq10}), however, which 
is {\it independent} of the particular choice of transducers in an
interferometer and constitutes a genuine quantum property of the
bipartite system, contains {\it three} equally real but mutually
exclusive quantities. 
Comparing Eqs.\ (\ref{eq10}) and (\ref{eq13}), we can identify the
missing  quantity in Refs.\ \cite{jaeger1,jaeger2} as the
predictability ${\cal{P}}_{k}$. The most general complementarity
relation for a bipartite system exposed to an interferometer reads
therefore as 
\begin{equation}
{\cal{V}}_{12}^{2}+{\cal{V}}_{k}^{2}+{\cal{P}}_{k}^{2}=1. 
\label{eq15}
\end{equation}
Consequently, there is a correspondence between 
the ``generalized'' visibility $V_{k}$ in (\ref{eq14}) and 
the single-partite property ${\cal{S}}_{k}$. 
This correspondence is given by   
\begin{equation}
V_{k}^{2} \leftrightarrow {\cal{V}}_{k}^{2}+{\cal{P}}_{k}^{2}
={\cal{S}}_{k}^{2}.  
\label{eq16}
\end{equation} 
We stress that ${\cal{S}}_{k}$ is 
invariant under local unitary transformations 
and is, thus, independent of the explicit form 
of the transducers. 
Of course, the quantities ${\cal{P}}_{k}$ and 
${\cal{V}}_{k}$ explicitly depend on the form 
of the transducers involved. In fact, $V_{k}(={\cal{S}}_{k})$ is the
maximum of ${\cal{V}}_{k}$ under such transformations.  

Another important complementarity relation which is intrinsically 
related to bipartite systems was derived by Englert \cite{englert}
and, independently, by Jaeger {\it et al.} \cite{jaeger2}. 
This relation is given by 
\begin{equation}
{\cal{D}}^{2}+{\cal{V}}^{2} \leq 1, 
\label{eq17}
\end{equation} 
where ${\cal{D}}$ is the distinguishability and 
${\cal{V}}$ is the visibility. In case of pure bipartite quantum
systems the inequality transforms into an equality. 
The distinguishability is a measure of the {\it possible} 
which-path information in an interferometer that Nature 
can grant us. This possible information can be stored in an additional 
system which is {\it entangled} to the quantum system under 
consideration. 
Although the quantities involved were originally introduced as
apparently single-partite quantities, one of them, the 
distinguishability ${\cal{D}}$, is intrinsically  
related to a bipartite property because it depends on 
the {\it correlation} to the additional quantum system which 
serves as the possible information storage. 
Indeed, it was qualitatively recognized, that the 
distinguishability ${\cal{D}}$ should be related to 
an entanglement measure \cite{englert2}. 

With the new complementarity relation (\ref{eq10}) at hand, we can  
now find a relation between the distinguishability, ${\cal{D}}$, 
and the entanglement measure, ${\cal{C}}$. 
In order to do this, we explicitly label the quantities 
in Eq.\ (\ref{eq17}) with the system index $k=1,2$ 
of the bipartite system. Comparing Eqs.\ (\ref{eq10}) and
(\ref{eq17}), we readily find the relationship,
\begin{equation}
{\cal{D}}_{k}^{2}={\cal{C}}^{2}+{\cal{P}}_{k}^{2}. 
\label{eq18}
\end{equation}
between distinguishability and concurrence. This expression reveals
explicitly which portion of the 
distinguishability ${\cal{D}}_{k}$ of subsystem $k$ is of nonclassical
origin. Clearly, ${\cal{D}}_{k}$ is bounded from below
by the single-particle property predictability. In the case when
${\cal{C}}>0$, it becomes {\it possible} to obtain additional
which-path information in excess of the predictability, with
appropriate sorting of the two-particle data. 
In the context of the recent debate on how complementarity is
enforced \cite{englert3,wiseman1,wiseman3,duerr,haroche}, 
this relation is of fundamental importance. 
It explicitly states that complementarity is enforced by 
correlations and not by the Heisenberg uncertainty relations 
in the standard formalism. In order to relate complementarity to the
uncertainty relations, the latter must be extended to
entangled composite systems in order to include the nonclassical
correlations between the subsystems \cite{bjork,rempe,trifonov,luis}. 

In the following, we will establish the connection between the general
complementarity relation  
(\ref{eq10}) [or (\ref{eq15})] and a quantitative quantum erasure
relation which was derived recently \cite{englert2,jakob1}. 
A similar relation in the context of entanglement 
generation and quantum statistics was found in Refs.\ 
\cite{bose1,bose2}. 
The erasure relation quantifies the maximum possible 
fringe visibility, i.e.\ the coherence ${{c}}_{k}$, 
of subsystem $k$, which can be achieved by quantum erasure, 
in form of an inequality \cite{englert2}, 
\begin{equation}
{{c}}_{k}^{2}+{\cal{P}}_{k}^{2}\leq 1. 
\label{eq19}
\end{equation}
In case of a {\it pure} bipartite quantum system the 
inequality transforms into an equality. 
It was later shown, for a special class of bipartite 
systems, in which no single partite visibility was present, 
that the quantity ${{c}}_{k}$, although a single system property, 
is directly related to the entanglement measure
concurrence, ${\cal{C}}$ \cite{jakob1}.  
With the help of the general complementarity relation 
(\ref{eq10}) [or (\ref{eq15}) in case of interferometric considerations] 
we can express the coherence, $c_{k}$, for the most general 
case of bipartite two-level systems as 
\begin{equation}
c_{k}^{2}={\cal{C}}^{2}+{\cal{V}}_{k}^{2}
\equiv {\cal{V}}_{12}^{2}+{\cal{V}}_{k}^{2}. 
\label{eq20}
\end{equation} 
Equations (\ref{eq18}) and (\ref{eq20}) yield for the 
concurrence
\begin{equation}
{\cal{C}}=\sqrt{c_{k}^{2}-{\cal{V}}_{k}^{2}}=
\sqrt{{\cal{D}}_{k}^{2}-{\cal{P}}_{k}^{2}}, 
\label{eq21}
\end{equation}
indicating a relation between the 
distinguishability ${\cal{D}}_{k}$ and the coherence 
$c_{k}$ as well as their relation to the entanglement 
measure concurrence ${\cal{C}}$. 
We note that ${\cal{D}}_{k}\geq {\cal{P}}_{k}$ and  
$c_{k}\geq {\cal{V}}_{k}$, which is 
obvious from their definition \cite{englert2}.   
Relation (\ref{eq21}) also reveals the intimate quantum 
aspect of distinguishability  and coherence when they 
exceed their lower bounds.  
We further note that relation (\ref{eq21}) gives us 
the opportunity to employ measurable entities for the 
quantification of entanglement. Thus, entanglement becomes 
an operational concept. 

So far we have restricted our treatment to cases where the bipartite
system is described in a {\it pure} quantum state. 
The question necessarily arises as to what happens when the system 
initially contains some mixture. 
Because of the relations (\ref{eq17}) and (\ref{eq19}), 
proved in Refs.\ \cite{jakob1,englert2}, we obtain, together with Eqs.\ 
(\ref{eq18}) and (\ref{eq20}), the following inequality 
for the most general case of a mixed bipartite qubit system 
\begin{equation}
{\cal{C}}^{2}+{\cal{V}}^{2}_{k}+{\cal{P}}^{2}_{k} \leq 1. 
\label{eq22}
\end{equation}
This inequality can be qualitatively understood if we recall that 
concurrence and single-particle visibility in {\it mixed} bipartite
qubit systems are limited by the upper bound of 
the corresponding quantities for {\it pure} systems. 
We emphasize that for {\it mixed} bipartite systems 
the above complementarity relation {\it can not} directly 
be rewritten into the corresponding inequality 
involving two-particle visibility ${\cal{V}}_{12}$, as 
was possible for the {\it pure} systems case. 
The reason is that the direct relation between concurrence and 
two-particle visibility ceases to exist for mixed quantum systems. 
In particular, the two-particle visibility, as defined by Jeager {\it et al.} 
\cite{jaeger1,jaeger2}, can differ from zero in mixed 
bipartite quantum systems although the concurrence is zero. 
This suggests a modification of the definition of 
two-particle visibility in mixed bipartite quantum systems which we 
plan to discuss elsewhere. 

In the case of {\it pure} bipartite systems an interesting 
relation emerges from the general complementarity 
relation (\ref{eq10}) for possible violations of the Bell inequality 
\cite{Bell1,Bell2,shimony}. 
In particular, we can state that violations of the Bell inequality 
are enforced by the complementarity between bipartite and single 
particle properties (\ref{eq12}).  
This becomes clear when we recall the well known relation 
between the maximum possible violation of the Bell inequality 
${\cal{B}}$ and the concurrence ${\cal{C}}$, 
\begin{equation}
{\cal{B}}=2\sqrt{1+{\cal{C}}^{2}}, \label{eq23}
\end{equation}
which is valid for pure bipartite qubit systems \cite{gisin}. 
With Eq.\ (\ref{eq12}) we can rewrite (\ref{eq23}) as 
\begin{equation}
{\cal{B}}=2\sqrt{2-{\cal{S}}_{k}^{2}}, \label{eq24}
\end{equation}
where the single partite property ${\cal{S}}_{k}$ of subsystem $k$ 
is given by Eq.\ (\ref{eq11}). 
As violations of the Bell inequality occur for $2<|{\cal{B}}|\leq
2\sqrt{2}$, Eq.\ (\ref{eq24}) reveals that exclusion of single partite 
properties in bipartite systems leads to maximal violation of the Bell
inequality. In other words, the complementarity relation (\ref{eq10})
governs the classical vs. nonclassical properties of the bipartite
system. If the local single partite properties  
${\cal{S}}_{k}$ are mutually excluded by the maximum possible 
concurrence, ${\cal{C}}=1$, the local realities necessarily 
cease to exist and we expect maximum possible violation 
of the Bell inequality. Thus, we may claim that violations of the Bell
inequality are also governed by complementarity between single
particle and bipartite properties. 

In conclusion, complementarity in bipartite systems 
is a useful concept for the quantification of entanglement. 
We have shown that concurrence (or the equivalent 
entanglement of formation) naturally emerges as 
a complementary quantity to single partite 
properties in a bipartite qubit system. 
The single partite properties consist of 
local but mutually exclusive realities given by  
visibility and predictability, i.e.\ wave-particle duality. 
Thus, the complementarity relation in bipartite 
qubits contains three mutually exclusive quantities, 
i.e.\ a ``triality'' relation is emerging naturally. 
The concurrence, herein, is an explicit bipartite property, 
and does not have any classical counterpart. 
Moreover, we have shown that violation of the Bell 
inequality becomes a consequence 
of the complementarity between bipartite 
and single partite properties. For future direction of research in
this area it might be worth while to explore complementarity concepts 
in multi-partite multiport interferometers. They provide a realization
of multipartite qudit systems where the concept of entanglement is
still not well understood.  

This research was supported by the Office of Naval 
Research under Grant No.\ N00014-92J-1233 and by the European 
Union Research and Training Network COCOMO, Contract No.\ 
HPRN-CT-1999-00129. 
\bibliographystyle{unsrt}

\end{document}